# Atomistic misconception of current model for condensed matter evaporation and new formulation

**V. V. Semak**

**Applied Research Laboratory, The Pennsylvania State University, PA, USA**

**Abstract**

Even though the phenomenon of evaporation is omnipresent and has immense scientific and technological importance, the research effort to unveil its fundamentals remains inadequately low. As one particular consequence, the textbooks and educational courses are lacking detailed explanation of evaporation and its effects. In order to advance fundamental theory of evaporation and increase accuracy of evaporation simulation a novel evaporation theory is presented. This integrated Atomistic(Molecular)-Kinetics-Gasdynamics theoretical model that combines statistical mechanics, gas dynamics and thermodynamics approaches opens a path to detailed description of nonstationary, nonequilibrium evaporation of condensed matter. The main innovation of the proposed approach is that, unlike all previous and current models of evaporation that are based on the assumption of evaporation as emission of the particles that are not bound within the condense phase, the described new model treats evaporation of condensed phase as escape of the particles of sufficient kinetic energy out of potential well located at the boundary of condensed and gaseous phases. Correspondingly, the re-condensation of the vapor onto the surface is treated as entrapment of the vapor particles with kietic energy lower than the depth of the potential well. The described novel research will open new opportunities to substantially advance our knowledge and provide needed contributions to chemical, combustion, environmental, climate and other sciences that utilize evaporation theory. Additionally, it will provide invaluable new material for the undergraduate and graduate educational courses in Physics, Chemistry, Engineering, and Material Sciences.

## Subject of the article

Evaporation is perceived by general community of scientists, engineers and educators as well studied and it is commonly believed that accurate physical model is mostly completed. However, those who attempt practical application of the existing models find that accuracy of predictions is unacceptably low for any real-world application. This indicates that current understanding of seemingly trivial evaporation phenomenon is inadequate. To large extent, the lack of interest to achieving deeper understanding of evaporation is a consequence of recent general trend in research that demands rushing forward in pursue of short term benefit. History of science and technology shows that significant and rapid progress can be achieved in particular disciplines leaving unanswered deeper questions about fundamental nature of the universe. However, at certain stage of such “modular” development the further progress becomes impossible without re-examining these multiple fundamental questions left behind.

In this article we describe re-examination and further development of evaporation theory utilizing approach that combines atomistic statistical mechanics, gas dynamics and thermodynamics concepts. Novelty of the presented theoretical approach is that, our model treats evaporation of condensed phase as escape of the particles with sufficient kinetic energy from a potential well located at the boundary of condensed and gaseous phases. This is unlike all the previous and current models of evaporation that assume the evaporation as emission of the particles that are not bound within the condense phase.

## Background

Currently there are two theoretical approaches in study of evaporation: Classical Kinetic Theory [1-3] in combination with Continuum Mechanics Theory [4-6] and, recently proposed, Statistical Rate Theory (SRT) [7].

The SRT model of evaporation is based on quantum and statistical mechanics with complex math and it is still not verified and is in process of further development. Although so far, the predictions from the SRT were found to be in agreement with the measurements [7] its complexity and insufficient completeness prevents wide utilization for practical purposes.

The dominant and mostly researched theoretical model of evaporation (such as [4-6]) is based on the concept proposed more than hundred years ago by Hertz, Langmuir, and Knudsen [1-3] (referred here as HLK model). This theoretical model is a straightforward physical model expressed in terms of relatively simpler mathematics; however, the HLK model in current interpretation is known to lack accuracy expected for practical applications. Because of this low accuracy the HLK model is used only for the educational purposes: it is taught as a concept to the students studying Thermodynamics, Heat Transfer and other related disciplines. In practice, researchers and application engineers use empirical relations (when such are available) that provide the evaporation rate, saturated vapor pressure, and other vapor parameters as function of temperature, pressure and other conditions that seldom include velocity and chemical composition of ambient gas.

As an example of practical application, in the simulation of laser material interaction the evaporation is very important and during past several decades an empirical evaporation model has been created. This model was successfully used in simulation of laser machining of materials [9-11] for the surface temperatures ranging from melting temperature to slightly below critical temperature. However, generation of the empirical data for evaporation modeling requires significant effort and thus, data are available for limited number of materials and even more limited environment conditions. The obtained equations contain adjustment coefficients with values spreading in wide range and, so far, the attempts to formulate first-principle approach for determination of these coefficients were unsuccessful.

The supposition proposed here is that low accuracy of the HLK model results from inadequacy of the foundational assumption of this model that at the evaporating surface the vapor particles have half-Maxwellian distribution function of the component of velocity normal to the condensed phase. This assumption intrinsically implies that the evaporation is an expansion of unbound and not interacting particles, i.e. the depth of potential well at the boundary between the condensed and gaseous phases is either neglected or assumed to be zero. This assumption, proposed more than 100 years ago when Statistical Mechanics and Solid State Physics were in infant form, remained unchallenged since, although it obviously contradicts current commonly accepted theoretical concept of condensed matter as collection of particles that are bound to each other resting within potential wells.

Consistently with the theoretical models of condensed phase we propose a new theory that describes evaporation as escape of the surface particles from this potential well. Thus, in order for a particle to cross the interface from condensed material into ambient gas its kinetic energy should exceed the depth of the potential well that is formed due to the particulates bonding. Below we will demonstrate of how this seemingly trivial conceptual modification lads to entirely different vapor equations that do not require empirical adjustment factors and are capable of providing broader insight into evaporation phenomenon and dramatic increase of simulation accuracy.

**Description of previous evaporation model**

The main assumptions of the HLK model are the following:

I. The evaporated particles have half-range Maxwellian distribution function $f_e$ for the velocity component normal to the evaporating surface, $\xi_y$, that is denoted as positive directed into the gas phase and negative directed into the condensed phase ($y$-axis is normal to the surface)

$$f_e(v_y) = \left(\frac{m}{2\pi k T_s}\right)^{1/2} \exp\left(-\frac{m\xi_y^{\,2}}{2kT_s}\right), \qquad \text{for } \xi_y > 0 \qquad (1)$$

$$f_e(\xi_y) = 0, \qquad \text{for } \xi_y \le 0$$

Where $m$ is the mass of the particle, $k$ is the Boltzmann's constant, and $T_s$ is the temperature at the condensed phase surface. Note: this assumption implicitly suggests that the evaporation is equivalent to the expansion of not bound particles, i.e. potential barrier of zero depth at the interface between condensed and gaseous phases.

II. After the vapor particles reach the edge of so called Knudsen layer with thickness O(λ) that is on the order of magnitude of mean free path length, $\lambda$, a full Maxwellian distribution function is established, $f_\infty$,

$$f_\infty(\vec{v}) = \left(\frac{m}{2\pi k T_v}\right)^{3/2} \exp\left(-\frac{m\left(\vec{\xi} - \vec{c}\right)^2}{2kT_v}\right), \qquad (2)$$

where $T_v$ is the temperature of vapor, and $\mathbf{\vec{c}}_\infty = (0, c_\infty, 0)$ is the one dimensional velocity vector on the outer edge of Knudsen layer, i.e. vapor flow gasdynamic velocity.

Further improvement of HLK model was performed in the past 50 years [4-6], and the follow-up modified model used assumption that, the distribution function of the evaporated material has form

$$f_\infty(y,\xi) = a_+^e(y)\, f_e^+(\xi) + a_\infty^+(y)\, f_\infty^+(\xi) + a_\infty^-(y)\, f_\infty^-(\xi), \qquad (3)$$

here $f_e^+$ is the $\xi_y > 0$ half-range function given in equation (1) and $f_\infty^+$ and $f_\infty^-$ are similar half-range functions defined by the equation (2), and the $a$-coefficients are empirical "adjustment" coefficients.

Thus, practical use of this interpretation of the HLK model requires experimental measurement of the ***a***-coefficients for evaporation and condensation. After the empirical coefficients are obtained for the given conditions, one should use so called moment method [4-6] consisting in solving the equations for conservation of mass, momentum and energy in order to determined parameters of vapor at the outer edge of Knudsen layer ( $\rho_\infty, T_\infty, v_\infty$ ) as functions of the material surface temperature, ambient pressure and other environmental parameters.

As an example of such use of HLK model, the rate of mass loos from unit area in unit time due to evaporation-condnsation is given by the following equation [8]

$$\dot{m}_{ev} = \sigma_e m n_w \left(\frac{m}{2\pi k T_w}\right)^{1/2} \int_0^\infty v_y exp\left(-\frac{m v_y^2}{2kT_w}\right) dv_y - \sigma_c m n_\infty \left(\frac{m}{2\pi k T_\infty}\right)^{1/2} \int_{-\infty}^0 v_y exp\left(-\frac{m v_y^2}{2kT_\infty}\right) dv_y, \quad (4)$$

where $n_w$ and $T_w$ are number density and temperature of vapor near the wall, $\boldsymbol{n}_\infty$ and $\boldsymbol{T}_\infty$ are the number density and temperature of vapor at large distance from the evaporating surface, $\sigma_e$ and $\sigma_c$ are evaporation and condensation coefficients, correspondingly. The evaporation and condensation coefficients in equation (4) are equivalent to the *a*-coefficients in the equation (3) and these coefficients are determined from the experiments (further description of these coefficients can be found in [6-8]).

If the gasdynamic velocity of vapor, *c*, is assumed to be negligibly small and the vapor is assumed to be an ideal gas (i.e. $P=nkT$), the integration of distribution functions given by equations (1,2) substituted into equation (4) provides well known and widely used Schrage's formula [4]:

$$\dot{m}_{ev} = \left(\frac{m}{2\pi k}\right)^{1/2} \left[\frac{\sigma_e}{1-0.5\sigma_e}\frac{P_w}{T_w^{1/2}} - \frac{\sigma_c}{1-0.5\sigma_c}\left(\frac{m}{2\pi k}\right)^{1/2}\frac{P_\infty}{T_\infty^{1/2}}\right]. \quad (5)$$

Typically [7], it is assumed that vapor pressure at the wall, $P_w$, equals the saturated vapor pressure, $P_{sv}$, at the temperature of the surface, $T_s$, i.e.

$$\dot{m}_{ev} = \left(\frac{m}{2\pi k}\right)^{1/2} \left[\frac{\sigma_e}{1-0.5\sigma_e}\frac{P_{sv}(T_s)}{T_s^{1/2}} - \frac{\sigma_c}{1-0.5\sigma_c}\left(\frac{m}{2\pi k}\right)^{1/2}\frac{P_\infty}{T_\infty^{1/2}}\right]. \quad (6)$$

In order to compute the net rate of material evaporation in addition to empirically determined coefficients of evaporation and condensation one has to know pressure at large distance from the evaporating surface, $P_\infty$. This requires solving mixed kinetic – gasdynamic equations known as "moment equations" [6]. It is typically suggested that the conditions are close to evaporation in vacuum and then, the condensation (second) term in the equation (6) can be omitted and the material loss rate is determined by the surface temperature and corresponding saturated vapor pressure – the evaporation (first) term in the equation (6).

The saturated vapor condition by definition is the condition when evaporation and condensation rates are equal and the net mass loss from the surface of condensed phase is zero. The pressure of vapor under condition of saturation is typically expressed using Clausius-Clapeyron equation

$$\frac{dP_{sv}(T)}{dT} = \frac{L_v}{T\Delta V}, \qquad (7)$$

where $L_v$ is the specific latent heat for evaporation, $T$ is the temperature of surface and vapor that equal since thermal equilibrium is assumed, and the $\Delta V$ is the change of material's specific volume during transition from condensed to vapor phase. Currently, there is no first principles theoretical model that provides temperature dependencies of the specific latent heat of evaporation and change of material's specific volume during phase transition. Typically it is assumed that the vapor is an ideal gas and the specific volume of gas phase, $V^g$, is much larger than specific volume of condensed phase, $V^c$, (the latter is acceptable if the temperature is much lower than the critical temperature). Then

$$\Delta V = V^g - V^c \approx V^g \approx \frac{RT}{P_{sv}} , \qquad (8)$$

where $R$ is the universal gas constant. Following earlier proposed approximation (formula (4.90 in [12]), the temperature dependence of the specific latent heat of evaporation can be expressed with simple algebraic equation

$$L_v = A + BT + CT^2, \qquad (9)$$

where $A$, $B$, and $C$ are empirical constants. On substituting equations (8) and (9) into (7) and integrating

$$\ln P_{sv} = -\frac{A}{RT} + \frac{B}{R}\ln T + \frac{C}{R}T + D, \qquad (10)$$

where D is the constant of integration.

Another frequently used empirical expression for the temperature dependence of saturated vapor pressure is Antoine equation [13]

$$\ln P_{sv} = A - \frac{B}{C+T}, \qquad (11)$$

where, similarly, $A$, $B$, and $C$ are substance specific empirically determined coefficients.

In a narrow range of temperatures the latent heat of evaporation can be assumed as temperature independent and then simplified Clausius-Clapeyron equation can be obtained following integration of equation (7) assuming validity of approximation (8)

$$P_{sv} = A\exp\left(-\frac{L_v}{RT}\right), \qquad (12)$$

where $A$ is a constant of integration that must be determined empirically, for example, using measured temperature of boiling under normal conditions and corresponding value of saturated vapor pressure that equals to 1atm.

**Description of new theoretical model**

In order to advance fundamental theory of evaporation, eliminate need for empirical equations, and to increase accuracy of evaporation simulation we propose a novel evaporation theory that combines statistical mechanics, gas dynamics and thermodynamics approaches. This integrated Atomistic(Molecular)-Kinetics-Gasdynaics theoretical model is based on first-principle concepts and provides consistent description of nonstationary and nonequilibrium evaporation of condensed matter using fundamental physical properties of material. In particular, this new model is capable of coherently describe evaporation into ambient gaseous media of dissimilar nature and variable pressure in wide span of temperatures ranging up to the critical temperature.

The main innovation of the proposed conceptual approach is that, unlike previous models of evaporation based on the assumption of evaporation as emission of the particles that are not bound within the condense phase, the described here new model treats evaporation of condensed phase as escape (re-adsorption) of the particles of sufficient kinetic energy from (into) a potential well located at the boundary of condensed and gaseous phases.

The foundational assumptions of our model are the following:

I. The particles of condensed phase have Maxwellian distribution function and the emitted particles have velocity vector directed away from the surface of condensed phase and kinetic energy exceeding depth of the potential well, $U_0$ (Figure 1, top);

II. After several collisions the particles of evaporated material establish a new Maxvellian distribution function that corresponds to a gasdynamic flow of vapor with velocity $c$. The evaporated particles that have normal to the surface velocity component directed toward the surface and by amplitude exceeding gas flow velocity, $c$, will be re-adsorbed (condense) onto the surface if their kinetic energy is lower than the depth of the potential well, $U_0$ (Figure 1, bottom).

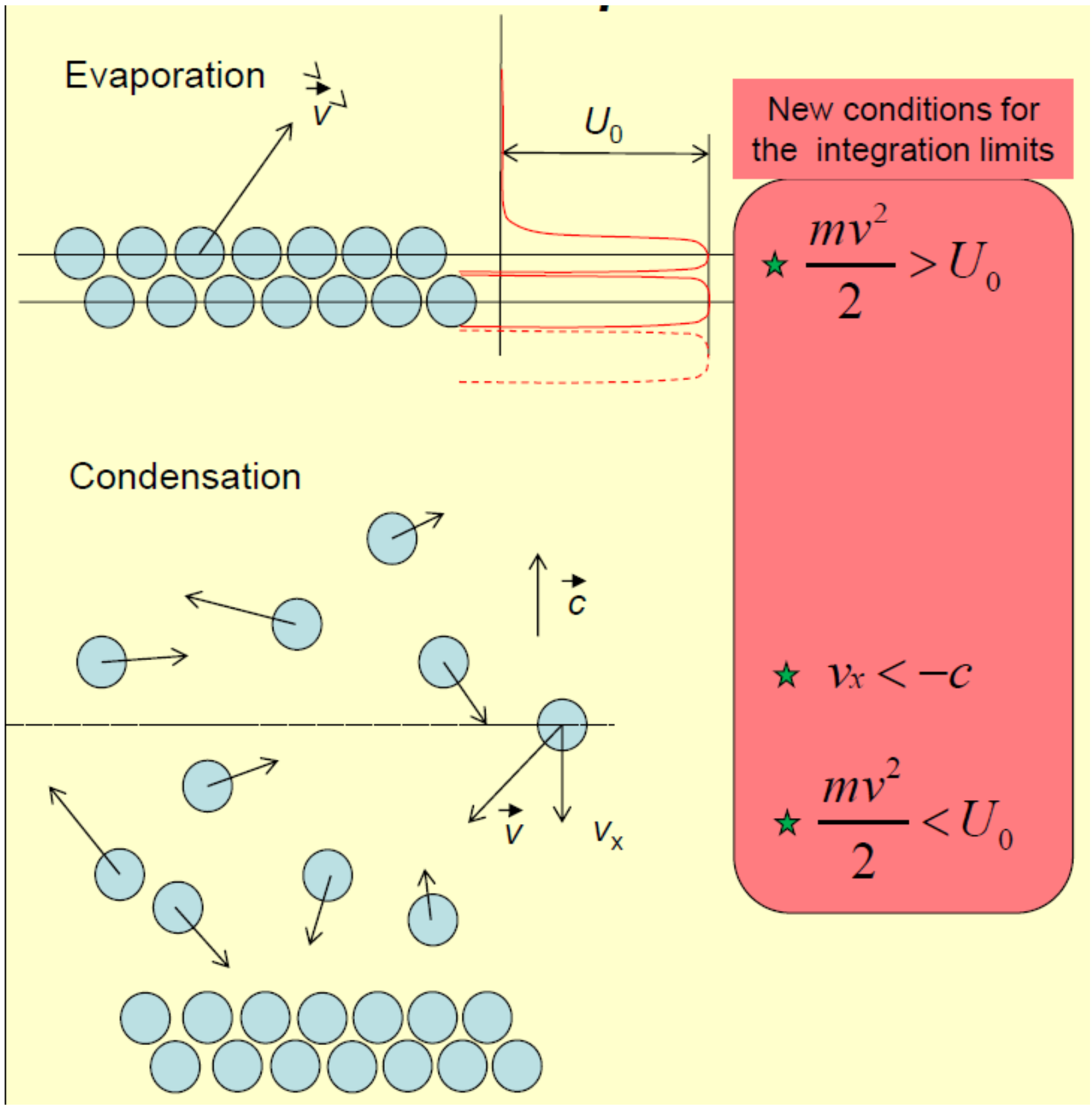

Figure 1. Schematic representation of new theoretical model for surface evaporation and re-condensation onto the surface.

Integrating the Maxwellian distribution function at the evaporating surface while applying the integration velocity limits that satisfy the condition for escaping from the potential barrier (Figure 1, top), i.e. normal component of velocity is positive (the particle moves out and away from the condensed phase) and the kinetic energy higher than the depth of the potential barrier, and equating this integral to the gasdynamic flow rate gives the equation for conservation of mass, momentum and energy (for evaporation in vacuum or low ambient pressure):

$$\mu_e = n_s \int d\vec{v} v_x f_e(\vec{v}) = n_v c\,, \qquad (13)$$

$$\omega_e = n_s \int d\vec{v} v_x \vec{v} f_e(\vec{v}) = \frac{P_v}{m} + n_v c^2\,, \qquad (14$$

$$\varepsilon_e = n_s \int d\vec{v} v_x \vec{v}^2 f_e(\vec{v}) = n_v c\left(c^2 + 3\frac{kT_v}{m}\right), \qquad (15)$$

here the subscript “e” is for evaporation, $n_s$ is the number density in the condensed phase, $c$ is the vapor flow gas-dynamic velocity, $P_v$ is the pressure of vapor (partial pressure) and $T_v$ is the vapor temperature at the end of the Knudsen layer.

Solving the system of three equations (13-15) with three unknowns gives three macroscopic parameters of the vapor at the end of Knudsen layer: $n_v$, $c$, and $T_v$ (the vapor pressure is expressed via vapor density and temperature - $P_v = kn_sT_v$).

Knowing the macroscopic parameters of vapor at the end of Knudsen layer provides the vapor velocity distribution function in the coordinate system of the material surface

$$f_c(\vec{v}) = \left(\frac{m}{2\pi k T_v}\right)^{3/2} \exp\left(-\frac{m(v-c)^2}{2kT_v}\right), \qquad (16)$$

and thus, allows computing (without need for any empirical coefficients) the mass, momentum and energy fluxes due to the condensation:

$$\mu_c = n_v \int d\vec{v} v_x f_c(\vec{v}), \qquad (17)$$

$$\omega_c = n_v \int d\vec{v} v_x \vec{v} f_c(\vec{v}), \qquad (18)$$

$$\varepsilon_c = n_v \int d\vec{v} v_x \vec{v}^2 f_c(\vec{v}), \qquad (19)$$

where the subscript "c" is for condensation and the integration of the distribution function (16) is performed according to the limits shown in the Figure 1, bottom.

Seemingly, integration of equations (13-15,17-19) can be attained in analytical form and, as an example, we will present integration of equations (13) and (17) for evaporation and condensation mass fluxes, correspondingly. Under the assumptions of our model the left side of equation (13) representing evaporation mass flux can be written in the following form:

$$\mu_e = n_s \int d\vec{v} v_x f_e(\vec{v}) = n_s \left(\frac{m}{2\pi k T_s}\right)^{3/2} \int_{v_{\min}}^{\infty} v^3 \exp\left(-\frac{mv^2}{2kT_s}\right) dv \int_0^{\pi/2} \cos\theta \sin\theta d\theta \int_0^{2\pi} d\varphi, \qquad (20)$$

where $n_s$ is the number density of the condensed phase, $m$ is the mass of the evaporated particle (atomic or molecular), $T_s$ is the surface temperature, $v$ is the modulus of the particle velocity vector, angles $\theta$ and $\phi$ are angles of velocity vector in the plane normal to and coinciding with plane of the surface of condensed phase, correspondingly. The limits of integration for velocity follow from the requirement that the kinetic energy of the evaporating particle should exceed the depth of potential well at the material surface, $U_0$, and, therefore, $v > v_{\min} = \sqrt{\frac{2U_0}{m}}$. The integration limits for the angles reflect that the evaporation flux contain particles moving into a half-space.

Using variables substitution, $x = \frac{mv^2}{2kT_s}, y = \cos\theta$, and integrating over angle $\phi$ allows rewriting equation (20) in the following form

$$\mu_e = \frac{1}{2} n_s \left(\frac{m}{2\pi k T_s}\right)^{3/2} \left(\frac{2kT_s}{m}\right)^2 2\pi \int_{x_{\min}}^{\infty} x \exp(-x) dx \int_{1}^{0} -y dy =$$
$$\frac{1}{2} n_s \left(\frac{2\pi k T_s}{m}\right)^{1/2} \left[-\exp(-x)(1+x)\right]_{x_{\min}}^{\infty} = \frac{1}{2} n_s \left(\frac{2\pi k T_s}{m}\right)^{1/2} \exp\left(-\frac{U_0}{kT_s}\right)\left(1+\frac{U_0}{kT_s}\right) . \quad (21)$$

Next, we integrate left side of equation (17) representing the condensation mass flux, i.e. flux of the vapor particulates that directed toward the condensed phase surface and have component of the velocity normal to the phase surface that is larger than the gasdynamic velocity of vapor flow, *c*, and with kinetic energy that is smaller than the depth of potential well at the material surface, $U_0$. For convenience, we perform integration in the coordinate system of moving vapor assuming positive direction toward the condensed phase surface and then the left side of the equation (17) is as follows

$$\mu_c = n_v \int d\vec{v} v_x f_c(\vec{v}) = n_v \left(\frac{m}{2\pi k T_v}\right)^{3/2} \int_{v_{\min}}^{v_{\max}} \int_{0}^{\theta_{\max}} v^3 \exp\left(-\frac{mv^2}{2kT_v}\right) \cos\theta \sin\theta dv d\theta \int_{0}^{2\pi} d\varphi . \quad (22)$$

Because the variables $v$ and $\theta$ are not independent the separation of variables similar to equation (20) is impossible; however, the variable $\phi$ is independent and the separate integration over this variable is allowed. Making substitution of variables, $x = \frac{mv^2}{2kT_v}$ and $y = \cos\theta$, and integrating over $\phi$ gives

$$\mu_c = \frac{1}{2} n_v \left(\frac{m}{2\pi k T_v}\right)^{3/2} \left(\frac{2kT_v}{m}\right)^2 2\pi \int_{x_{\min}}^{x_{\max}} \int_{0}^{\theta_{\max}} x \exp(-x) \cos\theta \sin\theta dx d\theta =$$
$$n_v \left(\frac{2\pi k T_v}{m}\right)^{1/2} \int_{x_{\min}}^{x_{\max}} \int_{y_{\min}=\cos(0)}^{y_{\max}=\cos(\theta_{\max})} -x \exp(-x) y dx dy . \quad (23)$$

Now we compute the integration limits in the equation (23) using mentioned above conditions necessary for the back flow of vapor to condense on the surface:

1) $$v_x = v\cos\theta > c \Rightarrow v > \frac{c}{\cos\theta} \quad (24)$$

2) $$\frac{m\vec{v}_r^{\,2}}{2_s} = \frac{m(\vec{v}-\vec{c})^2}{2} < U_0 . \quad (25)$$

From simple geometry $|\vec{v}_r|^2 = |\vec{v}|^2 + |\vec{c}|^2 - 2|\vec{v}||\vec{c}|\cos\theta$ and then from (25) it follows

$$v^2 - 2vc\cos\theta + c^2 - \frac{2U_0}{m} < 0 . \quad (26)$$

The inequality (26) is satisfied when

$$c\cos\theta - \sqrt{c^2(\cos^2\theta - 1) + \frac{2U_0}{m}} < v < c\cos\theta + \sqrt{c^2(\cos^2\theta - 1) + \frac{2U_0}{m}}\,, \qquad (27)$$

and taking into consideration that right side of inequality (24) is larger than left side inequality (25) we finally arrive to the following inequality that defines the integration limits for velocity

$$v_{\min} \equiv \frac{c}{\cos\theta} < v < c\cos\theta + \sqrt{c^2(\cos^2\theta - 1) + \frac{2U_0}{m}} \equiv v_{\max}\,. \qquad (28)$$

Thus, from inequality (28) taking into account substitution of variable $y = \cos\theta$, the limits for integration over variable $x$ in the integral (23) are

$$x_{\min} = \frac{m}{2kT_v}\frac{c^2}{\cos^2\theta} = \frac{m}{2kT_v}\frac{c^2}{y^2}\,, \qquad (29)$$

and, assuming that $c << \sqrt{\frac{2U_0}{m}}$,

$$x_{\max} = \frac{m}{2kT_v}\left(c\cos\theta + \sqrt{c^2(\cos^2\theta - 1) + \frac{2U_0}{m}}\right)^2 \approx \frac{U_0}{kT_v}\,. \qquad (30)$$

The upper integration limit $\theta_{\max}$ can be determined from equating left and right sides of inequality (28)

$$\frac{c}{\cos\theta_{\max}} = c\cos\theta_m + \sqrt{c^2(\cos^2\theta_{\max} - 1) + \frac{2U_0}{m}}\,. \qquad (31)$$

Since $c^2 << \frac{2U_0}{m}$ then the first term under the square root sign can be neglected leading to simplification of the equation (31)

$$c\cos^2\theta_{\max} + \sqrt{\frac{2U_0}{m}}\cos\theta_{\max} - c = 0\,. \qquad (32)$$

The solutions of equation (32) are

$$\cos\theta_{\max\,1,2} = \frac{-\sqrt{\frac{2U_0}{m}} \pm \sqrt{\frac{2U_0}{m} + 4c^2}}{2c}\,. \qquad (33)$$

Only positive solution is physical and, with simplification we have

$$\cos\theta_{\max} = \frac{\sqrt{\frac{2U_0}{m}}\left(\sqrt{1+{\frac{4c^2}{2U_0/m}}^{2}}-1\right)}{2c} \approx \sqrt{\frac{2U_0}{m}}\,\frac{c}{2U_0/m} = \frac{c}{\sqrt{\frac{2U_0}{m}}}, \tag{34}$$

and, finally,

$$\theta_{\max} = \arccos\left(\frac{c}{\sqrt{\frac{2U_0}{m}}}\right). \tag{35}$$

Thus, from equation (35) the limits for integration over variable $y$ in the integral (23) are

$$y_{\min} = 1, \tag{36}$$

$$y_{\max} = \frac{c}{\sqrt{\frac{2U_0}{m}}}. \tag{37}$$

Using the integration limits for $x$ (29,30) and for $y$ (36,37) we can integrate the equation (23)

$$\mu_c = n_v\left(\frac{2\pi kT_v}{m}\right)^{1/2}\int_{x_{\min}}^{x_{\max}}\int_{y_{\min}=\cos(0)}^{y_{\max}=\cos(\theta_{\max})} -x\exp(-x)ydxdy =$$

$$n_v\left(\frac{2\pi kT_v}{m}\right)^{1/2}\int_{x_{\min}=\frac{m}{2kT_v}\frac{c^2}{y^2}}^{x_{\max}=U_0/kT_v}\int_{y_{\min}=1}^{y_{\max}=c/\sqrt{2U_0/m}} -x\exp(-x)ydxdy =$$

$$n_v\left(\frac{2\pi kT_v}{m}\right)^{1/2}\int_{y_{\min}=1}^{y_{\max}=c/\sqrt{2U_0/m}} \exp(-x)(1+x)\Big|_{\frac{m}{2kT_v}\frac{c^2}{y^2}}^{U_0/kT_v} ydy =$$

$$n_v\left(\frac{2\pi kT_v}{m}\right)^{1/2}\int_{y_{\min}=1}^{y_{\max}=c/\sqrt{2U_0/m}} \left[\exp\left(-U_0/kT_v\right)\left(1+U_0/kT_v\right)-\exp\left(-\frac{m}{2kT_v}\frac{c^2}{y^2}\right)\left(1+\frac{m}{2kT_v}\frac{c^2}{y^2}\right)\right]ydy =$$

$$n_v\left(\frac{2\pi kT_v}{m}\right)^{1/2}\left[\begin{array}{l}\exp\left(-U_0/kT_v\right)\left(1+U_0/kT_v\right)\frac{1}{2}y^2\Big|_{y_{\min}=1}^{y_{\max}=c/\sqrt{2U_0/m}} - \\ \int_{y_{\min}=1}^{y_{\max}=c/\sqrt{2U_0/m}}\exp\left(-\frac{m}{2kT_v}\frac{c^2}{y^2}\right)\left(1+\frac{m}{2kT_v}\frac{c^2}{y^2}\right)\frac{1}{2}d(y^2)\end{array}\right] =$$

$$\frac{1}{2}n_v\left(\frac{2\pi kT_v}{m}\right)^{1/2}\left[\begin{array}{l}\exp\left(-U_0/kT_v\right)\left(1+U_0/kT_v\right)\left(\frac{mc^2}{2U_0}-1\right)- \\ \frac{mc^2}{2kT_v}\int_{z_{\min}=\frac{2kT_v}{mc^2}}^{z_{\max}=\frac{kT_v}{U_0}}\exp\left(-\frac{1}{z}\right)\left(1+\frac{1}{z}\right)dz\end{array}\right] =$$

$$\frac{1}{2}n_v\left(\frac{2\pi kT_v}{m}\right)^{1/2}\left[\begin{array}{l}\exp\left(-U_0/kT_v\right)\left(1+U_0/kT_v\right)\left(\frac{mc^2}{2U_0}-1\right)- \\ \frac{mc^2}{2kT_v}z\exp\left(-\frac{1}{z}\right)\Big|_{z_{\min}=\frac{2kT_v}{mc^2}}^{z_{\max}=\frac{kT_v}{U_0}}\end{array}\right] \quad ,(38)$$

where we introduce variable substitution $z = y^2\frac{2kT_v}{mc^2}$, such that $z_{\min} = \frac{2kT_v}{mc^2}$ and $z_{\max} = \frac{2kT_v}{mc^2}\frac{mc^2}{2U_0} = \frac{kT_v}{U_0}$. Thus, finally the mass flux of evaporated material that condenses back onto the surface is

$$\mu_c = \frac{1}{2} n_v \left( \frac{2\pi k T_v}{m} \right)^{1/2} \left[ \begin{array}{l} \exp\left( -U_0 / kT_v \right)\left( 1 + U_0 / kT_v \right)\left( \dfrac{mc^2}{2U_0} - 1 \right) - \\ \left( \dfrac{mc^2}{2U_0} \exp\left( -U_0 / kT_v \right) - \exp\left( -mc^2 / 2kT_v \right) \right) \end{array} \right]. \quad (39)$$

In the similar manner the equations for conservation of momentum and energy of the evaporated material (14,15) and the flux of momentum and energy back to the evaporating surface due to condensation (18,19) were analytically integrated and the results will be presented in the following publications.

In the future presented new theoretical approach can be further developed to describe evaporation into ambient gas of similar or dissimilar nature, with any pressure or temperature and any flow field pattern and velocity by adding equations that describe gasdynammics and heat exchange.

**New equation for saturated vapor pressure**

Two equations (21) and (39) describing mass flows of evaporated and condensed material can be used in order to obtain the vapor pressure under saturation condition. Under this condition the mass flows of evaporated and condensed material are equal, the gasdynamic velocity of vapor is zero and the vapor temperature equals the temperature of condensed phase, i.e.

$$\begin{array}{l} \mu_e = \mu_c \\ T_s = T_v = T \,. \\ c = 0 \end{array} \quad (40)$$

Then we can express density of saturated vapor in terms of particle density in condensed media, $n_s$, its temperature, $T$, and the depth of potential well at the boundary of condensed phase, $U_0$,

$$n_v = n_s \frac{\exp\left( -U_0 / kT \right)\left( 1 + U_0 / kT \right)}{1 - \exp\left( -U_0 / kT \right)\left( 1 + U_0 / kT \right)} . \quad (41)$$

Assuming that the vapor is an ideal gas, the new expression for saturated vapor pressure as function of temperature is as follows

$$P_{sat} = n_s k T_s \frac{\exp\left( -\dfrac{U_0}{kT} \right)\left( 1 + \dfrac{U_0}{kT} \right)}{1 - \exp\left( -\dfrac{U_0}{kT} \right)\left( 1 + \dfrac{U_0}{kT} \right)} , \quad (42)$$

where $U_0$ is the depth of potential barrier at boundary between the condensed and vapor/gas phases, $n_s$ is the number density in the condensed phase, and $T$ is the surface temperature of condensed phase.

The new equation (42) for saturated vapor pressure significantly differs from the known Clausius-Clapeyron equation (12) not only in form but in its fundamental meaning. Indeed, in addition to the temperature dependence, the new dependence for saturated vapor pressure includes depth of potential well at the surface, $U_0$. This quantitative physical property of material reflects "strength" of the bond between the particulates of condensed phase at the boundary with gaseous phase. It can be expressed as a function of fundamental material properties and its value can be computed from the models and theories that belong in the realm of solid state physics. It is naturally expected that the depth of the potential well, $U_0$, should be a decreasing (but not vanishing to zero) function of surface temperature; however, to the best of our knowledge, there is no currently a theory that provides detailed description of the interaction of particulates comprising condensed matter at the boundary with gaseous phase.

In contrast, the traditional expression for saturated vapor pressure contains the specific latent heat of evaporation, $L_v$, that is defined as the enthalpy change required to transform a given quantity of a substance from a condensed phase (liquid) into a gaseous phase at a given pressure. It is known that the latent heat of evaporation is a quantitative physical property that depends on temperature. Typically, in the thermodynamics or heat transfer courses and in textbooks the nature of latent heat of evaporation is qualitatively described as reflecting the bond between the condensed phase particulates; however, no detailed explanation for this property is given beyond the general statements. The temperature dependence of latent heat of evaporation is presented as empirical fact without detailed explanation of the nature of such dependence. In particular, the peculiar decrease of latent heat of evaporation down to zero at the critical temperature is not commented upon or explained; although, curious students should arrive to a puzzling conclusion that the bonding between the matter particulates vanishes at temperatures equal or exceeding the critical temperature. The latter, of course, contradicts to common sense. Additionally, in the educational courses it is seldom mentioned that latent heat of evaporation depends on ambient pressure and, when it is mentioned, such dependence is also presented as an empirical fact without explanations.

The new theoretical model presented here uses logical concept of the depth of potential well at the surface of condensed phase and, in particular, it clearly explains the nature of the latent heat of evaporation and opens a rout to detailed explanation of its dependences on temperature and ambient conditions as due to complex interplay of kinetics of evaporation and condensation. Indeed, following our model one can find that the specific latent heat of evaporation can be expressed as

$$L_v = \frac{\varepsilon_e - \varepsilon_c}{\mu_e - \mu_c}, \qquad (43)$$

where the fluxes of energy and mass due to evaporation and condensation can be computed in a manner shown above. It is obvious from our model (see equations (13-15,17-19)) that, the latent heat of evaporation is a complex function of surface temperature and environment conditions including vapor and ambient gas temperature, pressure, flow velocity, chemical composition, etc.

It is worth mentioning here that typically values of latent heat of evaporation are measured for certain temperature (usually boiling) and normal atmospheric pressure. For few materials the temperature dependence is measured in a narrow temperature ranges and even for fewer materials the temperature range extends to the critical temperature. To a large extent the reason for this is that the measurements of the latent heat of evaporation, as well as evaporation and condensation rates, is a difficult procedure and further development of the presented model in combination with the measurements and theoretical studies of the bonding of condensed phase particulates will open wide possibilities for accurate numerical simulation of evaporation of various materials in wide range of temperatures.

## Saturated vapor pressure: computed vs measured

In order to demonstrate utility of the new evaporation theory we performed comparison of empirical data for saturated vapor pressure of iron, aluminum and lithium with the values computed using the new model under assumption of temperature independent depth of potential barrier $U_0$ and values computed using typical Hertz-Langmuir-Knudsen/Clausius-Clapeyron model under assumption of constant latent heat of evaporation $L_v$ shown in the Figure 2. The empirical data were taken from the monograph [12] and the data for latent heat of evaporation were taken from the Wikipedia websites for corresponding meals.

The results shown in Figure 2 demonstrate that for heavier metals, such as aluminum and iron, all curves are closely similar; however, for very light lithium metal our model coincides with experiment data while the HLK/Clausius-Clapeyron equation gives prediction that is orders of magnitude different from the measurement.

Thus, the presented results allow to conclude that for considered metals in the range of temperatures up to boing point the depth of potential well at the surface, $U_0$, remains practically constant. Also, the simplified Clausius-Clapeyron equation, in which the constant latent heat of evaporation measured at the boiling temperature is used, reasonably reproduces measured values of saturated vapor pressure for iron and aluminum. However, for lithium temperature dependence of latent heat of evaporation can't be ignored and use of simplified Clausius-Clapeyron equation leads to significant discrepancy.

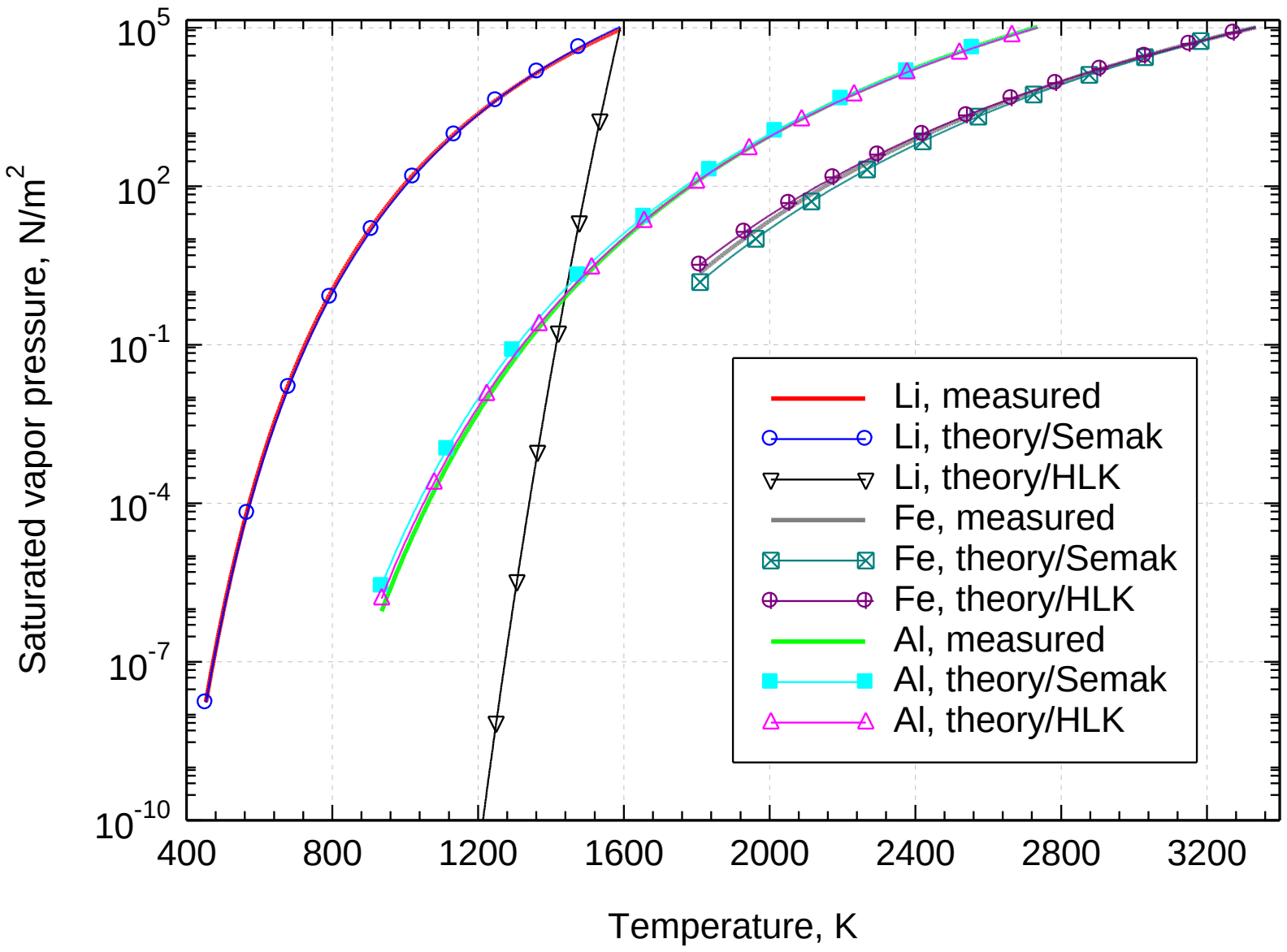

Figure 2. Metal saturated vapor pressure dependence on temperature: comparison of the new theory ("Semak") with simplified Hertz-Langmuir-Knudsen/Clausius-Clapeyron equation ("HLK") and experimental data ("measured").

We also tested our model for prediction of the saturated pressure of water. The values calculated using equation (42) demonstrated that for water a temperature dependent depth of potential battier, $U_0$, is required in order to achieve match between computed and experimental values. The temperature dependence of the depth of potential well at the surface expressed in J/mol (i.e., $U_0 N_a$, where $N_a$ is the Avogadro number) that provides best fit of our computations to the experimental data is shown in the Figure 3. The experimentally measured data for latent heat of evaporation of water as function of temperature is also shown in the Figure 3. There is notable difference between these two dependencies. The measured function $L_v(T)$ decreases to zero when temperature approaches critical temperature; however, the temperature function for the depth of potential well per mole, that provides match to the experimentally measured saturated vapor pressure, remains at a significantly above zero value at the critical temperature. The latter is consistent with concept that the bond between the particulates of condensed phase does not vanish at critical temperature and, in general, represent possible indirect way of measuring the bond properties in condensed phase and a method for verification of theoretical models of condensed matter.

Finally, a peculiar and seemingly unreported previously discrepancy should be pointed out. When the measured temperature dependent specific latent heat of evaporation of water, $L_v(T)$, shown in the Figure 3, is used in the Clausius-Clapeyron equation (7) the resulting computed values of saturated vapor pressure are very different from the experimentally measured values (Figure 4); however, one would reasonably expect complete match. This discrepancy illustrates and supports our supposition expressed above in discussion of the equation (43) for the latent heat of evaporation. Indeed, since the latent heat of evaporation is a complex function of many parameters that include ambient conditions, the measured values correspond to a specific environment in which the measurements are performed. Then, the difference in environment conditions for the saturated vapor pressure measurements will cause observed discrepancy.

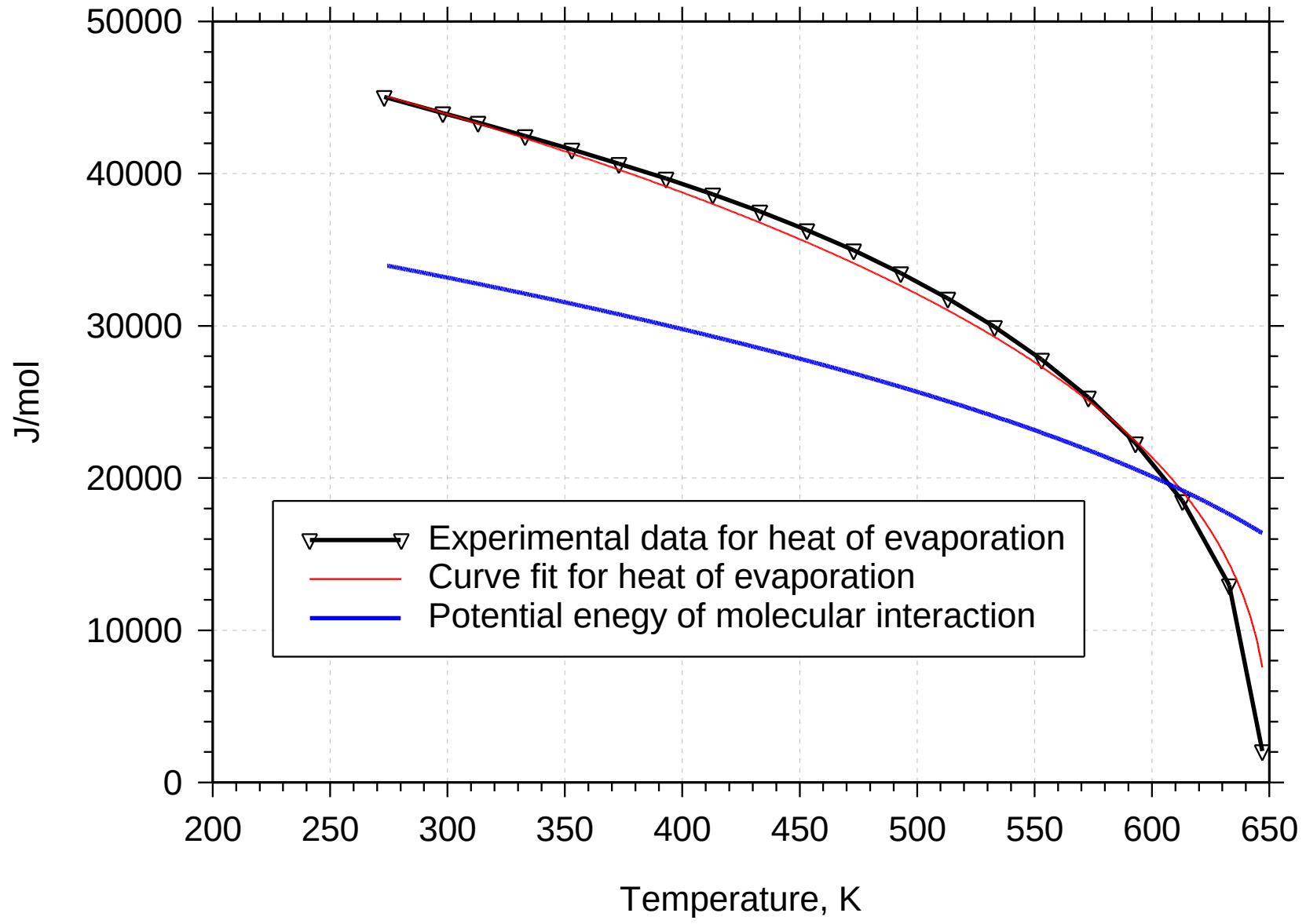

Figure 3. Experimentally measured specific latent heat of evaporation of water, $L_v$, (note it drops to zero at the critical temperature) and the product of temperature dependent depth of potential barrier, $U_0$, and the Avogadro number (in order to express potential well depth in units of J/mol).

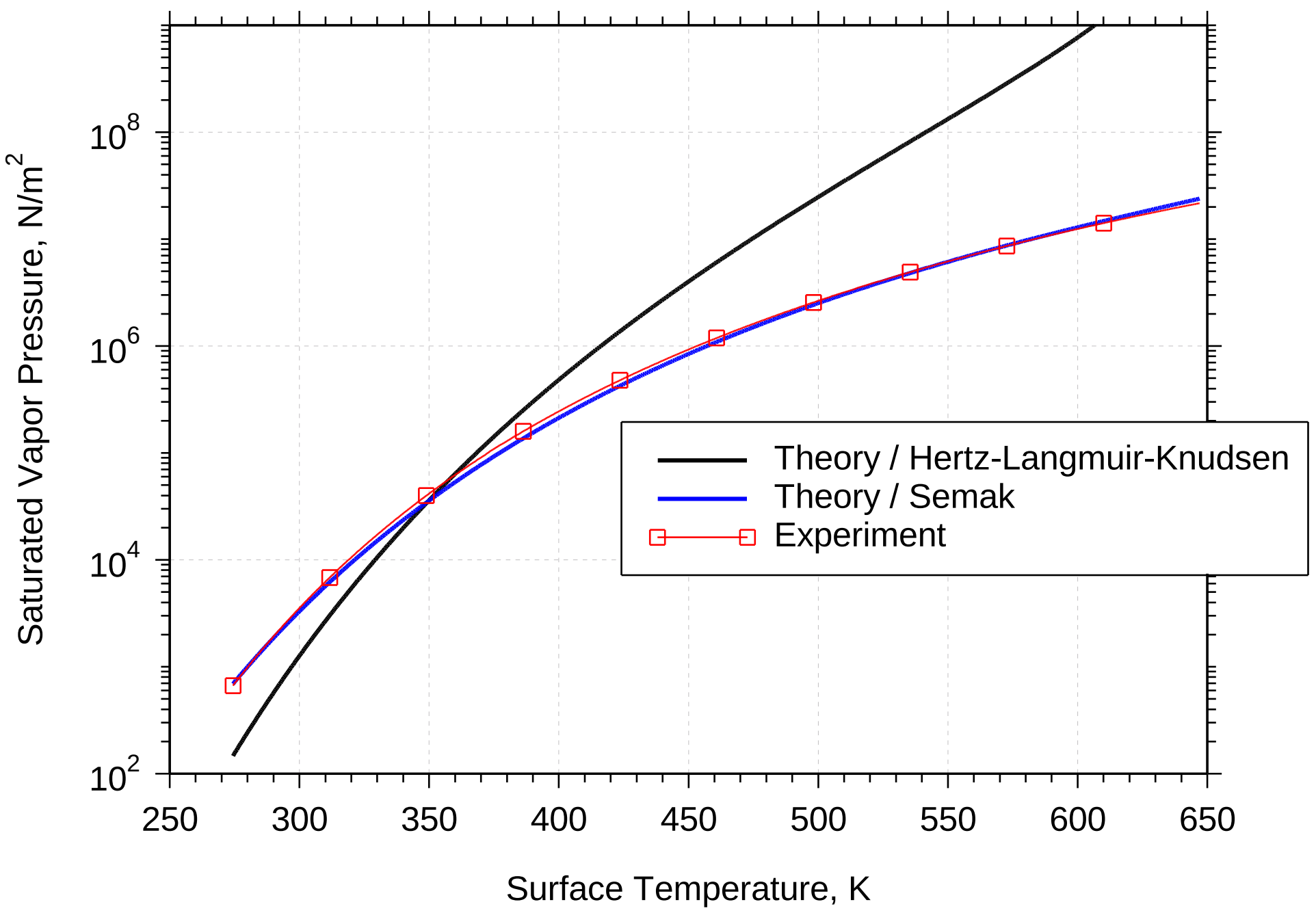

Figure 4. Water saturated vapor pressure dependence on temperature: comparison of the new theory ("Semak") with Hertz-Langmuir-Knudesen theory leading to Clausius-Clapeyron equation (HLK) and experimental data (experiment). The HLK curve was computed using measured temperature dependent latent heat of evaporation shown in Figure 3, and the new theory prediction was computed using temperature dependent depth of potential barrier shown in Figure 3 that gives best fit of saturated vapor dependence to experimental data.